\documentclass[%
 reprint,
superscriptaddress,
 amsmath,amssymb,
 aps,
]{revtex4-1}

\usepackage{graphicx}
\usepackage{dcolumn}
\usepackage{bm}

\begin{document}

\preprint{APS/123-QED}

\title{Stiffening solids with liquid inclusions}

\author{Robert W. Style}
\affiliation{%
Yale University, New Haven, CT 06520, USA 
}%
\author{Rostislav Boltyanskiy}%
\affiliation{%
Yale University, New Haven, CT 06520, USA 
}

\author{Benjamin Allen}%
\affiliation{%
Yale University, New Haven, CT 06520, USA 
}

\author{Katharine E. Jensen}%
\affiliation{%
Yale University, New Haven, CT 06520, USA 
}

\author{Henry P. Foote}%
\affiliation{%
University of North Carolina - Chapel Hill, NC 27516, USA
}

\author{John S. Wettlaufer}%
\affiliation{%
Yale University, New Haven, CT 06520, USA 
}
\affiliation{Mathematical Institute, University of Oxford, Oxford, OX1 3LB, UK}

\author{Eric R. Dufresne}%
\email[]{eric.dufresne@yale.edu}
\affiliation{%
Yale University, New Haven, CT 06520, USA 
}


%

\date{\today}

\begin{abstract}
From bone and wood to concrete and carbon fibre, composites  are ubiquitous natural and engineering materials.
Eshelby's inclusion theory describes how macroscopic stress fields couple to isolated microscopic inclusions, allowing prediction of a composite's bulk mechanical properties from a knowledge of its microstructure.
It has been extended to describe a wide variety of phenomena from solid fracture to cell adhesion.
Here, we show experimentally and theoretically that Eshelby's theory breaks down for small liquid inclusions in a soft solid.
In this limit, an isolated droplet's deformation is strongly size-dependent with the smallest droplets mimicking the behaviour of  solid inclusions.
Furthermore, in opposition to the predictions of conventional composite theory, we find that finite concentrations of small liquid inclusions enhance the stiffness of soft solids. 
A straight-forward extension of Eshelby's theory, accounting for the surface tension of the solid-liquid interface, explains our experimental observations.
The counterintuitive effect of liquid-stiffening of solids is expected whenever droplet radii are smaller than an elastocapillary length, given by the ratio of the surface tension to Young's modulus of the solid matrix.

\end{abstract}

\pacs{Valid PACS appear here}
\maketitle

\section{Introduction}
Composite materials can offer dramatic performance improvements over their individual components.
Carbon fibre increases the strength and stiffness of polymer resins as much as a hundredfold \cite{ashb89,baug02}.
Densely-packed gas bubbles in a liquid matrix create a foam, which resists deformation like a solid \cite{duri91,hohl05}.
The foundational theory of solid composites, due to Eshelby, describes how isolated inclusions in a composite behave in response to applied stresses \cite{eshe57}.
Eshelby applied this result to predict the stiffness of dilute solid composites \cite{eshe57} and his theory has been extended to finite concentrations, where neighbouring inclusions couple through their induced strain-fields (e.g. \cite{hash63,mori73,hill63,budi65}).  
Eshelby's theory has been applied widely beyond composites, having long been used to understand the mechanics of fracture \cite{rice68,budi76} and plasticity \cite{hutc70,berv78}. 
More recently, it has been applied to understanding flow of sheared glasses \cite{scha07} and the interactions of cells with the extracellular matrix \cite{zeme10,schw13}.

\begin{figure*}
\centering
\includegraphics[width=18cm]{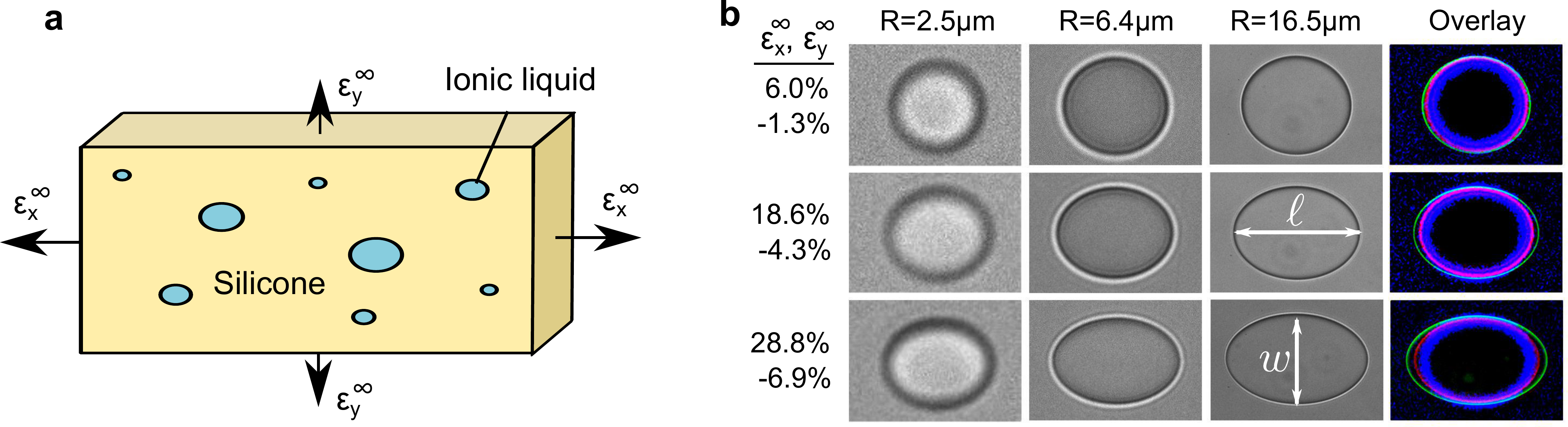}
  \caption{Stretching droplets embedded in soft solids. (a) The sample is clamped and stretched in the $x$-direction. (b) Example images of ionic-liquid droplets in a soft, silicone solid $E=1.7$kPa. Larger droplets deform more at the same applied strain. Overlay shows small (blue), medium (red) and large (green) droplet images combined together for shape comparison.}
  \label{fig:stretch_drops}
\end{figure*}

Eshelby's theory describes the matrix and inclusion as bulk linear-elastic solids, but does not account for physics of the interface, which generically includes excess surface free energy and surface stress.
Surface energy is the reversible work per unit area required to create new interfacial area by cutting.  
Surface stress is the reversible work per unit area to create new interfacial area by stretching.
For liquids, surface energy and surface stress are isotropic and identical.  
For solids, surface stress and energy are generally anisotropic and distinct, but can be isotropic for soft amorphous solids like gels \cite{shut50,spae00,hui13}.
Cell membranes and other thin-walled vessels can exhibit large isotropic surface stress with negligible surface energy \cite{need92,lecu07}.
In this manuscript, we use the phrase \emph{surface tension}, denoted by $\Upsilon$, to denote an isotropic surface stress.

Recent work has underlined the importance of surface tension effects in soft solids.
These solid capillary effects include the smoothing-out of ripples and corners in soft solids \cite{jago12,mora13}, and qualitative changes to the phenomena of wetting  \cite{styl12,styl13,styl13b,nade13,karp14} and adhesion  \cite{styl13c,sale13,xu14,cao14}.  
Additionally, the competition of surface tension and elasticity can select the wavelength of pearling and creasing instabilities \cite{mora10,chen12}.
Surface tension effects typically appear in solids at lengthscales $\lesssim L\equiv\Upsilon/E$, where $E$ is the Young modulus of the solid.
In simple terms, this elastocapillary length represents the wavelength below which surface tension is capable of significantly deforming a solid \cite{styl12,styl13c}.
Thus, it is reasonable to expect that when inclusions in an elastic body have a characteristic size $R<L$, capillarity will become important and Eshelby's theory will not apply. This has also been suggested in recent theoretical work (e.g. \cite{yang04,shar04,bris10}).

Here, we demonstrate the impact of surface tension on the mechanical response of fluid inclusions in a soft solid matrix.  
We find that the deformation of isolated liquid inclusions in a macroscopic stress field depends strongly on their size.   
While large droplet deformations are consistent with Eshelby theory, droplets with radii below the elastocapillary scale deform significantly less than predicted.
Furthermore, adding finite concentrations of large droplets to a solid makes it more compliant, while adding finite concentrations of droplets smaller than the elastocapillary scale makes it stiffer.  
A straight-forward generalisation of Eshelby's theory, accounting for surface tension, accounts for our experimental observations, and provides simple analytical results useful for the design of composites.

\section{Stretching single inclusions}

We tested Eshelby's inclusion theory in soft solids by observing the shape change of droplets embedded in soft/stiff solids as the solid is stretched (Figure \ref{fig:stretch_drops}a,b, Supplementary Information Section 1).
We coated the soft solid on a thin, elastic sheet, and stretched it uniaxially, measuring the exact applied strain by tracking fluorescent particles attached to the surface of the sheet (Supplementary Figure S1).
The applied strain consisted of a large tensile component along the stretch direction, $\epsilon_x^\infty$, and a small compressive strain perpendicular to this direction, $\epsilon_y^\infty$.
The stretch lengthens the droplets in the $x-$direction, and we imaged them at their equator from below with a 60x, NA 1.2, water objective.
The droplets are ionic liquid (1-ethyl-3-methylimidazolium dicyanamide, Ionic Liquids Technologies Inc.) and are completely immiscible in the silicone gel that we use for the solid phase.
Silicone gels of two different stiffnesses ($E=1.7$kPa, $100$kPa) were prepared by mixing together base and cross linker at different ratios, and curing at room temperature for 16 hours, as described in Section 2 of the Supplementary Information.
Silicone gel is ideal for these experiments as it behaves like a linear-elastic solid up to large strains.
Supplementary Figure S2 shows example rheology for the soft, $E=1.7$kPa silicone.

Eshelby predicts that stretched inclusion shapes depend only on the applied strain, and not on droplet size.
We confirmed this result for droplets embedded in a stiff, 100kPa matrix (Supplementary Figure S3).  
However, micron-sized droplets behave quite differently in a compliant 1.7kPa matrix (Figure \ref{fig:stretch_drops}b).
Here, small droplets are significantly less deformed than large droplets under the same macroscopic strain.

\begin{figure}
\centering
\includegraphics[width=8.5cm]{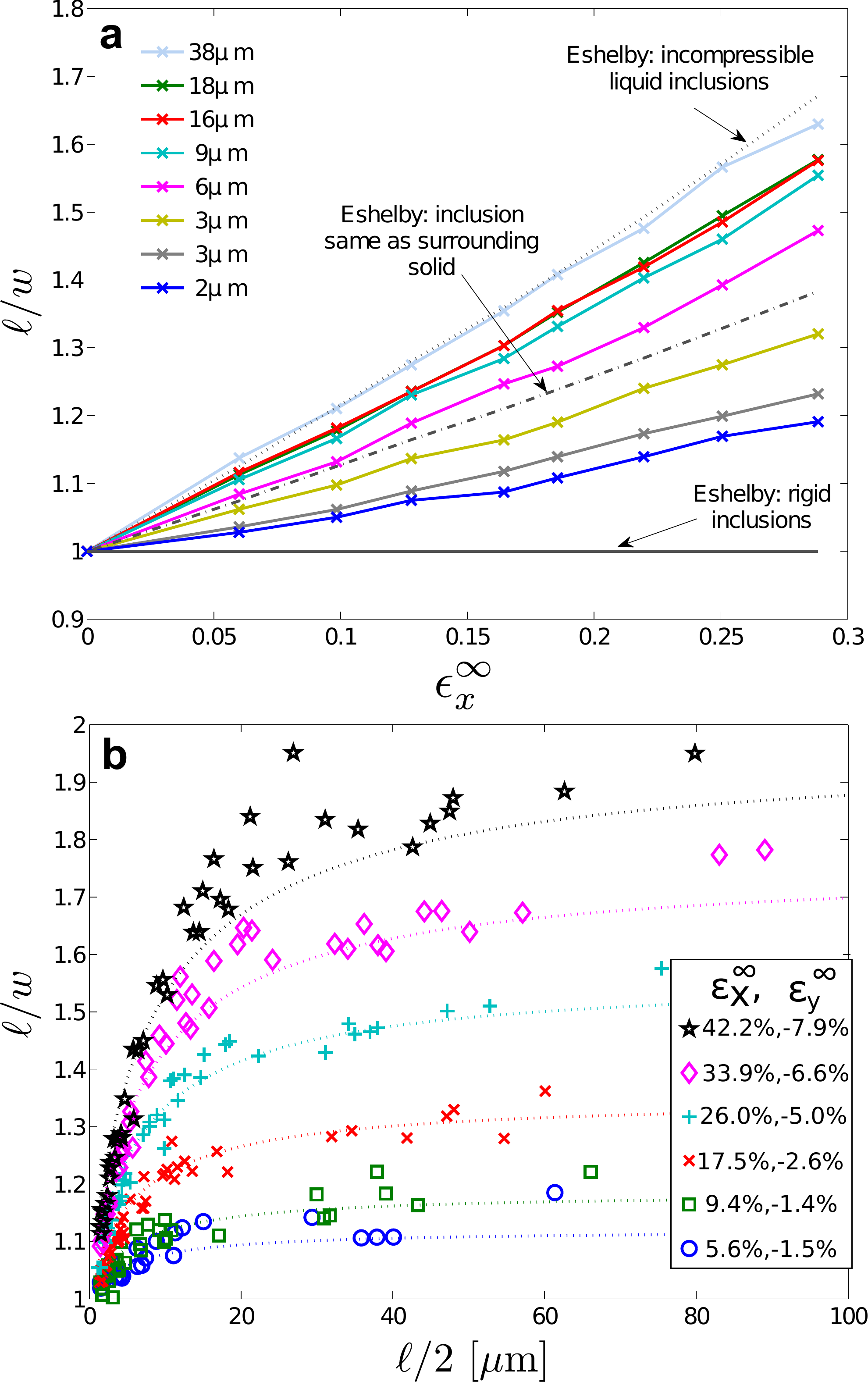}
  \caption{Aspect ratio of stretched ionic-liquid droplets in a soft ($E=1.7$kPa) silicone gel as a function of size and strain (a) Aspect ratios of 8 droplets of different sizes increase linearly with applied strain. The dashed-dotted line shows the change in aspect ratio of the solid under the applied strain $\epsilon_x^\infty,\epsilon_y^\infty$. Large droplets ($R\gtrsim 5\mu$m) stretch more than the solid, Smaller droplets stretch less, effectively stiffening the surrounding solid. The dotted/continuous lines show the predictions from Eshelby theory for incompressible liquid/rigid solid inclusions respectively \cite{eshe57}. (b) Aspect ratio of droplets depends sensitively on size. Different colours correspond to different applied strains. Dashed curves show theoretical predictions using Equations (\ref{eqn:l},\ref{eqn:w}) with $\Upsilon=0.0036$N/m.}
  \label{fig:data_drops}
\end{figure}

Small liquid droplets appear stiffer than the surrounding solid matrix.
Figure \ref{fig:data_drops}(a) gives the aspect ratio, $AR=\ell/w$, of 8 droplets of different initial radii $R$ at different stretches.
As expected for a linear elastic solid, the aspect ratio increases linearly with applied strain $\epsilon_x^\infty$.
However, it also increases with $R$.
In other words,  large droplets are deformed more by the stretch, and smaller droplets appear `stiffer'.
The dotted/continuous line shows Eshelby's predictions: $AR=(3+5\epsilon_x^\infty)/(3+5\epsilon_y^\infty)$ for incompressible, spherical liquid inclusions, and $AR=1$ for rigid spherical inclusions \cite{eshe57}.
We also plot Eshelby's prediction for an inclusion identical to the surrounding solid, $AR=(1+\epsilon_x^\infty)/(1+\epsilon_y^\infty)$ as the dashed line -- this represents the bulk deformation of the solid matrix.
The largest droplet agrees well with the incompressible liquid limit.
Smaller droplets appear stiffer, with the three smallest droplets deforming less than the solid matrix, and approaching the rigid inclusion limit.

The stiffening effect in small droplets appears to arise at a strain-independent lengthscale.
Figure \ref{fig:data_drops}(b) shows the aspect ratio of many droplets as a function of their length, $\ell$, for six different strains.
Again, this increases with strain, and small droplets appear stiffer than large droplets.
Aspect ratio is roughly constant for droplets $\gtrsim 30\mu$m, where it agrees relatively well with Eshelby's theory (not shown).
However $AR$ drops off sharply for smaller droplets.

\section{Composites that stiffen with increasing liquid content}

According to Eshelby's classic result \cite{eshe57}, liquid inclusions, which have zero Young's modulus, should reduce the stiffness of a solid composite.
However, our data shows that small, isolated droplets resist deformation more strongly than one would expect from Eshelby's theory.
Here we explore the impact of the increase in apparent stiffness of single droplets on the macroscopic stiffness of a composite.
We made soft composites out of silicone gel and glycerol droplets and measure the composite Young's modulus $E_c$ (see Supplementary Information Section 2 for detailed protocols).
We blended together silicone, glycerol (Sigma-Aldrich) and small quantity of surfactant (Gransurf 50C-HM, Grant Industries) with a hand blender.
Glycerol is used in place of the ionic liquid as it is cheap, non-toxic, and almost completely immiscible in silicone.
We  degassed the resulting emulsion in a vacuum, poured it into a mould and then cured it at 60$^\circ$C for two hours.
This gives composites of droplets embedded in silicone with $R\sim O(1\mu\mathrm{m})$, at volume fractions $\phi$ from 4 to 20\% (Supplementary Figure S4).
As explained in Supplementary Information Section 2b, we ignore composites with $\phi<4.4\%$ to ensure that the stiffness of the continuous phase of the composite is unaffected by the surfactant.  

\begin{figure}
\centering
\includegraphics[width=8.5cm]{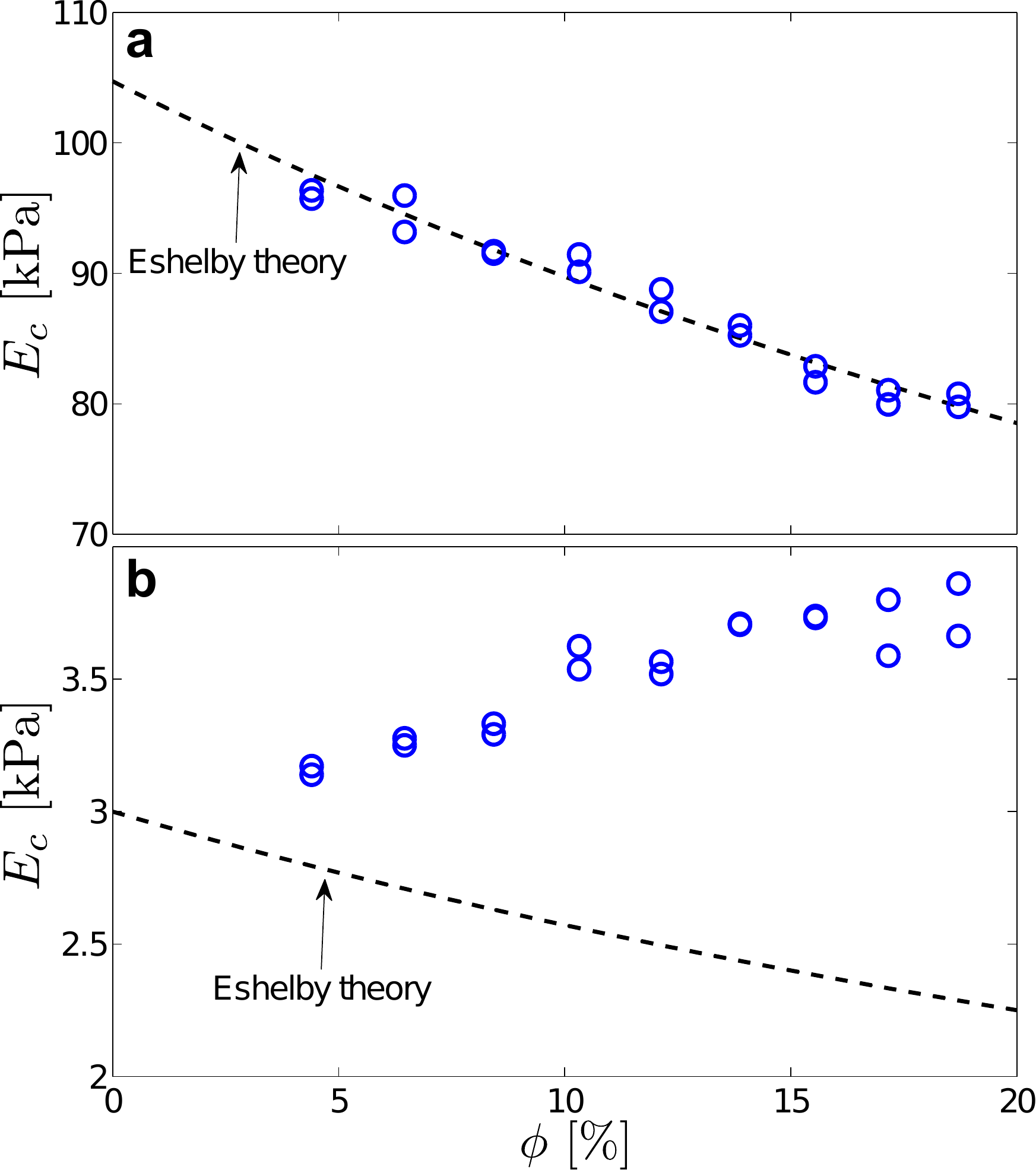}
  \caption{Young's modulus of soft composites as a function of liquid content. (a,b) Glycerol droplets embedded in $E\sim 3,100$kPa silicone gels respectively. Dashed curves show Eshelby's predictions for incompressible liquid droplets in an incompressible solid of stiffness $E$.}
  \label{fig:composites}
\end{figure}

We measured the stiffness of the composites by indenting a sample surface with a cylindrical metal rod of radius $a=1.6$mm, using an Instron with a 5N load cell.
The sample consisted of a filled (10mm deep $\times$ 35 mm diameter) petri dish.
From Hertz's law, the sample indentation, $d$, is related to the force applied by $F=8aE_c d/3$. 
Here we have assumed that the composite is incompressible as both silicone gel and glycerol are effectively incompressible (e.g. \cite{styl14}).
Thus we extract $E_c$ from the slope of the (initially-linear) force-displacement indentation profile (See Supplementary Figure S5 for example profiles). 

Stiff and compliant solids have opposing responses to liquid inclusions.  
Figure \ref{fig:composites} shows how composite stiffness changes with increasing liquid content for composites with (a) a stiffer solid matrix with $E\sim$100kPa and (b) a much more compliant solid matrix with $E\sim 3$kPa.
The stiff-matrix composite becomes softer as liquid content increases.
This makes intuitive sense -- as we replace a fraction of the solid by holes with no shear modulus, we see a proportional decrease to the stiffness.
In fact, the data agrees with Eshelby's prediction for the stiffness of a solid containing dilute embedded monodisperse, incompressible droplets, $E_c=E/(1+5\phi/3)$ \cite{eshe57}.
The compliant-matrix composite shows the opposite trend and  \emph{stiffens} with liquid content.
Stiffness increases by around a third with a $20\%$ increase in liquid content.
The composites are elastic up to shear strains of $\sim 100\%$, and behave identically in subsequent cycles of indentation (See Supplementary Information Section 2c and Supplementary Figure S5).
Thus, we find that the soft-matrix composite is unexpectedly stiffer than the pure soft solid without a significant loss in strength.
Conventional composite theory, such as Eshelby theory \cite{eshe57}, the Law of mixtures, and the Hashin-Shtrikman bounds \cite{hash63} uniformly predict decreasing stiffness with increased fraction of liquid inclusions and therefore cannot  describe this behaviour (e.g. Figure \ref{fig:composites}).

\section{Theory and Discussion}

The experimental data suggests that conventional composite theory fails to describe our experiments because of the effect of surface tension at the liquid/solid interface.
Surface tension typically acts to smooth out interfaces and drive them toward a constant curvature.
In a solid, surface tension is opposed by bulk elasticity.
However, surface tension can cause significant deformations in compliant solids (e.g. \cite{jago12,chak13}). 
In our experiments, surface tension acts to keep liquid inclusions spherical, opposing any applied stretch.
Thus, surface tension can qualitatively explain the main features of our data.
This  is  supported by recent experiments on wetting and adhesion on compliant silicones where capillary affects arose below a lengthscale of $O(10\mu$m$)$, similar to that seen in Figure \ref{fig:data_drops}(b) \cite{styl13,styl13b,styl13c}.
Here, we modify Eshelby theory to account for solid surface tension, and show that it accurately describes our data.

We consider an incompressible droplet embedded in a linear-elastic solid with a surface tension that acts on the droplet boundary.
In the solid, displacements, $\mathbf{u}$ obey the equation:
\begin{equation}
\label{eqn:elasticity}
(1-2\nu)\nabla^2\mathbf{u}+ \nabla(\nabla\cdot\mathbf{u})=0,
\end{equation}
where $\nu$ is Poisson's ratio and we apply far-field strain boundary conditions $\epsilon=\epsilon^\infty$.
At the surface of the droplet, $\sigma\cdot\mathbf{n}=-p+\Upsilon {\cal K}\mathbf{n}$, where $\sigma$ is the stress tensor in the solid, $\mathbf{n}$ is the normal vector to the deformed surface, $p$ is the pressure in the droplet and ${\cal K}$ is the curvature of the deformed surface.
Note that we assume that this surface tension is independent of surface strain. This is generally a good approximation for gels, though it may not be true in general \cite{shar04,hui13}.
We derive analytic solutions to these equations in Supplementary Information Section 3 by extending previous work \cite{duan05}.
In particular, for far-field, plane-stress boundary conditions (as in our experiment) $\epsilon_{xx}=\epsilon_x^\infty$, $\epsilon_{yy}=\epsilon_y^\infty$ and $\sigma_{zz}=0$, the length and width of the stretched droplet are
\begin{equation}
\label{eqn:l}
\ell=2R\left[1+\frac{5(2\epsilon_1-\epsilon_2)}{6+15\frac{\Upsilon}{ER}}\right],
\end{equation}
and
\begin{equation}
\label{eqn:w}
w=2R\left[1+\frac{5(2\epsilon_2-\epsilon_1)}{6+15\frac{\Upsilon}{ER}}\right],
\end{equation}
where $\epsilon_1=(\epsilon_x^\infty+\nu\epsilon_y^\infty)/(1-\nu^2)$ and $\epsilon_2=(\nu\epsilon_x^\infty+\epsilon_y^\infty)/(1-\nu^2)$.
In the limit $\Upsilon=0$, this reduces to Eshelby's predictions.
In the limit $\Upsilon/ER \gg 1$, surface tension dominates and the droplets stay spherical, as the elastic stresses becoming insufficient to deform the droplet from its preferred shape.   
The dependence on the parameter $\Upsilon/ER$ indicates that surface tension effects start to arise when droplets are similar in size to the elastocapillary length $L=\Upsilon/E$.
This is similar to previous experiments where solid capillarity becomes important: for example contact mechanics results are altered when the size of the indenter is $\lesssim L$ \cite{styl13c,sale13,xu14}, droplet contact angles change when drop radii are $\lesssim L$ \cite{styl12,styl13}, and thin fibres undergo instabilities when their diameters are $\lesssim L$ \cite{mora10}.

Our theory agrees well with the isolated droplet data with one fitting parameter -- the unknown surface tension $\Upsilon$.
In figure \ref{fig:data_drops}B, we plot the aspect ratio predicted by Equations (\ref{eqn:l},\ref{eqn:w}), using $\nu=1/2$ \cite{jeri11}, $E=1.7kPa$ and $\Upsilon=0.0036N/m$.
The results agree with the experiments up to surprisingly large strains, suggesting that surface tension is indeed controlling droplet shape for small droplets.
Note that the value of the surface tension is smaller than we expected;  
we measured surface tension of ionic liquid in uncured silicone to be 0.025N/m using the pendant drop method, and we might expect this value to be close to $\Upsilon$.
However, previous measurements have shown that there can be significant differences between liquid, and solid surface tensions of silicone \cite{styl13,jago12,nade13,park13}.

\begin{figure}
\centering
\includegraphics[width=9cm]{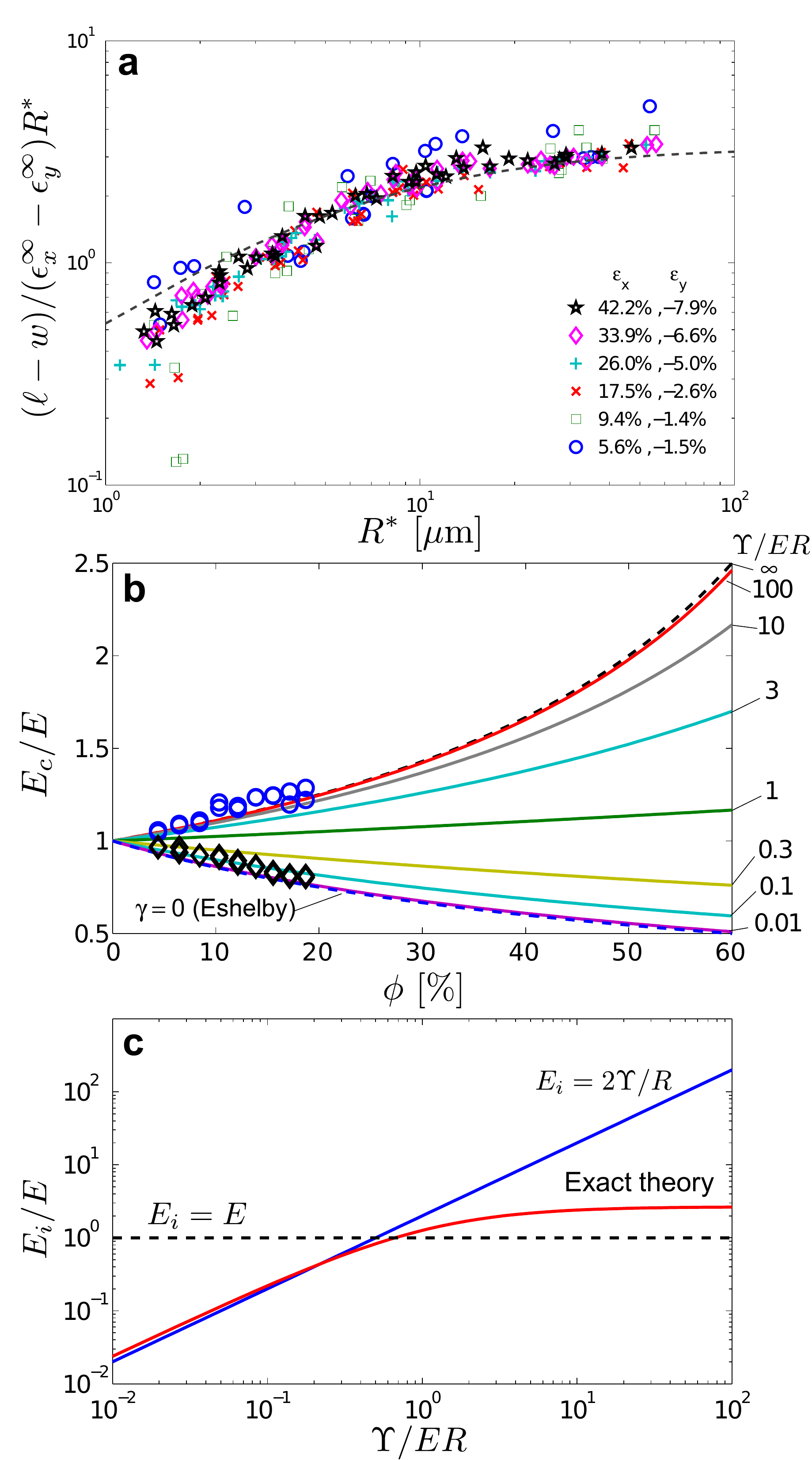}
  \caption{Theoretical predictions of composite behaviour. (a) The data from Figure \ref{fig:data_drops}(b) collapses when plotting estimated radius (Equation \ref{eqn:R}) against $2A/R^*$, a scaled measure of droplet sphericity (Equation \ref{eqn:A}). The dashed curve shows the theoretical prediction in Equation (\ref{eqn:R}). (b) Effective Young's modulus of a composite consisting of monodisperse droplets embedded in a uniform solid, from equation (\ref{eqn:E_c}). The blue/black data points are the composite stiffness data from Figure \ref{fig:composites} scaled by $E=3,100$kPa for the soft/stiff composites respectively. (c) Droplets with surface tension can be considered as equivalent elastic inclusions without surface tension. The red curve shows how the stiffness of the equivalent elastic inclusion, $E_i$ depends on $\Upsilon/ER$. For small $\Upsilon/ER$, this agrees quite well with the approximation $E_i=2\Upsilon/R$, shown by the blue line.}
  \label{fig:theory}
\end{figure}

Our elastocapillary theory collapses the experimental data collapses over a factor of 70 in droplet size, and a range of strains from $5.6-42.2\%$.  
From equations (\ref{eqn:l},\ref{eqn:w}), we can define the parameter
\begin{equation}
\label{eqn:A}
A\equiv\frac{(\ell-w)(1+\nu)}{3(\epsilon_x^\infty-\epsilon_y^\infty)}=\frac{10ER^2}{6ER +15\Upsilon}.
\end{equation}
$A$ can be thought of as a scaled measure of the noncircularity of the droplets. Using equations (\ref{eqn:l},\ref{eqn:w}), we then obtain an estimate of the droplet undeformed radii, $R^*$ ($R$ is unknown, as we did not track individual droplets from their undeformed state for this large data set).
\begin{equation}
\label{eqn:R}
R^*=\frac{\ell}{2}-\frac{A}{2}(2\epsilon_1-\epsilon_2).
\end{equation}
$A$ and $R^*$ only depend on measured quantities, and nicely collapse the data when plotted against each other (Figure \ref{fig:theory}a).
There are two regimes: for droplets of size $R^*\lesssim 10\mu$m, $(\ell-w)/R^* \propto (\epsilon_x-\epsilon_y)R^*$, while for larger droplets $(\ell-w)/R^*\propto (\epsilon_x-\epsilon_y)$. From equation (\ref{eqn:A}), we can interpret this as the crossover from capillary-dominated stretching when $R\ll\Upsilon/E$ to pure elastic stretching with $R\gg\Upsilon/E$ (where Eshelby theory applies).
Note that while our theory effectively collapses the data onto a universal curve, the data in the capillary regime appears to have a stronger dependence on droplet size than predicted.

The isolated droplet theory can also be used predict how the composite stiffnesses depend on liquid content \cite{eshe57,duan07a}.
Eshelby showed that the stiffness of a composite consisting of identical dilute inclusions can be calculated from the excess energy of individual strained inclusions \cite{eshe57};
if the extra strain energy due to the presence of a single inclusion in a uniaxially stretched solid is $W(\sigma^\infty, E, R, \Upsilon)$, where $\sigma^\infty$ is the applied stress, then the average strain energy density in a dilute composite is 
\begin{equation}
{\cal E}=\frac{1}{2}\frac{{\sigma^\infty}^2}{E}+\frac{\phi W}{\frac{4}{3}\pi R^3},
\end{equation}
and Young's modulus of the the composite is $E_c={\sigma^\infty}^2/2{\cal E}$.
In the Supplementary Material, we use this approach to predict the stiffness of a composite with monodisperse, incompressible inclusions with surface tension.
For the particular case of an incompressible solid,
\begin{equation}
\label{eqn:E_c}
E_{c}=E\frac{1+\frac{5}{2}\frac{\Upsilon}{ER}}{\frac{5}{2}\frac{\Upsilon}{ER}(1-\phi)+(1+\frac{5}{3}\phi)}.
\end{equation}
In the limit of small surface tension, or large droplets ($R\gg \Upsilon/E$), this reduces to Eshelby's result for liquid droplets in an elastic solid, $E_c=E/(1+5\phi/3)$.
When surface tension dominates elasticity ($R\gg \Upsilon/E$), we obtain $E_c=E/(1-\phi)$, and the material is stiffened by the inclusions.
This is not equivalent to Eshelby's result for rigid particles embedded in an elastic composite, $E_c=E/(1-5\phi/2)$ -- although surface tension keeps the droplets spherical, there is no shear stress at their surfaces, unlike the case of the rigid particles.
Equation (\ref{eqn:E_c}) also shows that composites are stiffened by droplets when $R<1.5\Upsilon/E$.
Figure \ref{fig:theory}(b) shows how composite stiffness depends on liquid fraction, as predicted by equation (\ref{eqn:E_c}) for different values of $\Upsilon/ER$.
We see the stiffening and softening described qualitatively above.
In particular, we see that the surface-tension-dominated regime is reached for $\Upsilon/ER\gtrsim 100$, and the Eshelby limit is observed for $\Upsilon/ER\lesssim 0.01$.

The theory for composite stiffness is consistent with our experimental data.
Figure \ref{fig:theory}(b) includes the data from Figure \ref{fig:composites}, normalised by $E=3,100$kPa for the softer/stiffer composites respectively.
The soft-matrix composite results are modelled well by the surface-tension dominated theory.
The stiff-matrix composite results are modelled well by the theory with little, or no, surface tension effects.
Using rough estimates of $\Upsilon\sim 14$mN/m (see Supplementary Information Section 2b) and $R\sim1\mu$m, we indeed expect surface tension to dominate for the soft-matrix composite ($\Upsilon/ER>1$), and to be small for the stiff-matrix composite ($\Upsilon/ER<1$).
Note that the experimental data for the soft-matrix composite is consistently stiffer than the upper limit of our theory.
We suspect that this is due to formation of chain-like structures of droplets (see Supplementary Figure S4).
Our theory is strictly only valid in the limit of isolated droplets.

We can greatly simplify the above results to give a simple physical picture of the effect of surface tension in soft composites.
The stiffness of a composite of incompressible elastic inclusions with Young's modulus $E_i$ in a solid of modulus $E$, according to Eshelby \cite{eshe57}, is
\begin{equation}
E_{c}=E\frac{1+\frac{2}{3}\frac{E_i}{E}}{\left(\frac{2}{3}-\frac{5\phi}{3}\right)\frac{E_i}{E}+(1+\frac{5}{3}\phi)}.
\end{equation}
If we equate this with Equation (\ref{eqn:E_c}), we find that embedded droplets are equivalent to elastic inclusions \cite{duan07a} with stiffness  
\begin{equation}
\label{eqn:eff_stiffness}
E_i=E\frac{24\frac{\Upsilon}{ER}}{10+9\frac{\Upsilon}{ER}}.
\end{equation}
When $\Upsilon/ER\ll 1$, the droplets behave like inclusions with Young's modulus $E_i=12\Upsilon/5R$.
This value is close to the droplet LaPlace pressure $E_i=2\Upsilon/R$, which is often taken as its stiffness for describing the composite stiffness of emulsions and gels \cite{van88,dick12}.
In the capillary dominated regime, $\Upsilon/ER\gg 1$, the effective Young's modulus of the inclusions saturates at $E_i=8E/3$.
Thus the droplets cannot have an arbitrarily increasing stiffness as they get smaller, as the $E'=2\Upsilon/R$ ansatz suggests.
By replacing capillary-dominated inclusions with equivalent elastic inclusions in this manner (e.g. \cite{duan07a}), one could use established composite theory such as Mori-Tanaka homogenisation \cite{mori73} or self-consistent methods \cite{budi65,hill65} to predict denser composite stiffnesses.

\section{Conclusions}

Our experimental and theoretical results  show that surface tension can be important for soft composites consisting of a liquid phase embedded in a continuous solid phase.
We expect that surface tension will be important for solid/solid composites whenever $R\lesssim 10\Upsilon/E_1,10\Upsilon/E_2$, where $E_1,E_2$ are the stiffnesses of the two solids.
For compliant materials, such as gels with $E\sim O(\mathrm{kPa})$, capillarity needs to be addressed at scales of up to $O(100\mu\mathrm{m})$ \cite{mora10,ducl14}.
For stiffer materials, such as elastomers, biopolymers, and soft nanocomposites, with $E\sim O(\mathrm{MPa})$, capillarity needs to be addressed at scales of up to $O(100\mathrm{nm})$.
Capillary effects should negligible in structural materials such as glass and ceramics with $E\sim O(\mathrm{GPa})$.
We expect that our results should be of use in understanding the mechanical properties of soft tissues, especially in soft connective tissues.
For example, the cortical tension of fibroblasts  may have a larger impact on the bulk mechanical properties of a collageneous tissue than the fibroblasts's elastic moduli \cite{brow98}.
Our results complement new approaches to measuring mechanical forces within three dimensional tissues by quantification of the deformation of embedded liquid droplets \cite{camp14}.
Our theoretical results include simple  analytic expressions for individual droplet deformation and for the properties of the bulk composite that can be readily applied to the design of new materials.  These results highlight important limitations to the common assumption that the effective stiffness of a droplet is equivalent to its Laplace pressure.

\section{Acknowledgements}

We thank Juan Fernandez-Garcia for the ionic liquids, and Tom Kodger \& Roger Diebold for advice in preparing silicone. We also thank Jon Singer, Michael Rooks, Frans Spaepen, Shomeek Mukhopadhyay, Peter Howell and Alain Goriely for helpful conversations. We gratefully acknowledge funding from the National Science Foundation (CBET-1236086) to ERD, the Yale University Bateman Interdepartmental Postdoctoral Fellowship to RWS and  the John Simon Guggenheim Foundation, the Swedish Research Council, and a Royal Society Wolfson Research Merit Award to JSW.

\end{document}